\documentstyle{agile7th}

\input epsf.sty
\input epsfig.sty
\input psfig.sty

\def \AAP #1 #2 {{\em Astron. Astrophys.\/} {\bf #1}, #2}
\def \AAL #1 #2 {{\em Astron. Astrophys. Lett.\/} {\bf #1}, L#2}
\def \AAR #1 #2 {{\em Astron. Astrophys. Rev.\/} {\bf #1}, #2}
\def \AAS #1 #2 {{\em Astron. Astrophys. Suppl. Ser.\/} {\bf #1}, #2}
\def \AJ #1 #2 {{\em Astron. J.\/} {\bf #1}, #2}
\def \ANNREV #1 #2 {{\em Ann. Rev. Astron. Astrophys.\/} {\bf #1}, #2}
\def \APJ #1 #2 {{\em Astrophys. J.\/} {\bf #1}, #2}
\def \APJL #1 #2 {{\em Astrophys. J. Lett.\/} {\bf #1}, L#2}
\def \APJS #1 #2 {{\em Astrophys. J. Suppl.\/} {\bf #1}, #2}
\def \APSS #1 #2 {{\em Astrophys. Space Sci.\/} {\bf #1}, #2}
\def \ASR #1 #2 {{\em Adv. Space Res.\/} {\bf #1}, #2}
\def \BAIC #1 #2 {{\em Bull. Astron. Inst. Czechosl.\/} {\bf #1}, #2}
\def \JSQRT #1 #2 {{\em J. Quant. Spectrosc. Radiat. Transfer\/} {\bf #1}, #2}
\def \MN #1 #2 {{\em Mon. Not. R. Astr. Soc.\/} {\bf #1}, #2}
\def \MEM #1 #2 {{\em Mem. R. Astr. Soc.\/} {\bf #1}, #2}
\def \PLR #1 #2 {{\em Phys. Lett. Rev.\/} {\bf #1}, #2}
\def \PASJ #1 #2 {{\em Publ. Astron. Soc. Japan\/} {\bf #1}, #2}
\def \PASP #1 #2 {{\em Publ. Astr. Soc. Pacific\/} {\bf #1}, #2}
\def \NAT #1 #2 {{\em Nature\/} {\bf #1}, #2}
\def \SAIT #1 #2 {{\em Mem.\ Soc.\ Astron.\ It.\/} {\bf #1}, #2}
\def \MESS #1 #2 {{\em The Messenger\/} {\bf #1}, #2}
\def \ASTRNACH #1 #2 {{\em Astron. Nach.\/} {\bf #1}, #2}
\def \AGPSR #1 #2 {{\em ASI Special Publication\/} {\bf #1}, #2}
\begin{opening}

\title{The AGILE Science Alert System}
\author{M. Trifoglio$^{1}$, A. Bulgarelli$^{1}$, F. Gianotti$^{1}$, F. Fuschino$^{1}$, M. Marisaldi$^{1}$, M. Tavani$^{2}$, E. Del Monte$^{2}$, Y. Evangelista$^{2}$, F. Lazzarotto$^{2}$, S. Sabatini $^{2}$, 
F. Longo$^{3}$, E. Moretti$^{3}$, C. Pittori$^{4}$ , F. Verrecchia$^{4}$ on behalf of the AGILE Team} 
\institute{$^1$INAF/IASF-Bologna I-40129 Bologna, Italy\\
$^2$INAF/IASF Roma I-00133 Roma, Italy\\
$^3$INFN Trieste, I-34127 Trieste, Italy\\
$^4$ASI Science Data Center, I-00044 Frascati (Roma), Italy
}
\date{} 
\end{opening}

\begin{document}

\oddpagefooter{}{}{} 
\evenpagefooter{}{}{} 
\medskip  

\begin{abstract} 
The AGILE Science Alert System  has been developed to provide prompt processing 
of science data for detection and alerts on $\gamma$-ray galactic and extra galactic 
transients,
$\gamma$-ray bursts, X-ray bursts and other transients in the hard X-rays.
The system is distributed among the AGILE Data Center (ADC) of the Italian Space Agency (ASI), 
Frascati (Italy), and the AGILE Team Quick Look sites, located at INAF/IASF Bologna and INAF/IASF Roma. 
We present the Alert System architecture and performances in the first 2 years of operation of the AGILE payload.  
\end{abstract}

\medskip

\section{Introduction}

The AGILE Mission (Tavani et al. 2009), dedicated to high-energy astrophysics, is designed 
to detect and image photons in the 30 MeV-50 GeV and 15-45 keV energy bands. 
The Payload complement consists of the Gamma-Ray Imaging Detector (GRID), 
with a large field of view, optimal time resolution and good sensitivity, 
combined with the hard X-ray monitor, Super-AGILE (SA), and the CsI(Tl) Mini-Calorimeter (MCAL).
AGILE was successfully launched on 23 April 2007, and during the first two years 
of operations it performed observations with fixed pointings,  
called Observation Blocks (OB), of typical duration of {$\sim$}2 weeks. 
Due to a {$\sim$}550 km equatorial orbit with inclination angle {$\sim$} 2.5$^{\circ}$, 
AGILE data are down-linked every {$\sim$} 100 minutes to the ASI Malindi ground station. 
From Malindi, the data are relayed through the ASI multi-mission network infrastructure 
(ASINet) to the Satellite Control Centre at Fucino, and then to the ASI Science Data Center (ASDC) 
at Frascati which hosts the AGILE Data Center (ADC) \footnote {http://agile.asdc.asi.it}. 
At each downlink, the AGILE raw TM received at ADC is automatically archived and transformed in FITS format (LV1) 
through the AGILE Pre-Processing System (TMPPS) (Trifoglio et al. 2008). 
All AGILE-GRID LV1 data are routinely processed 
using the scientific data reduction and analysis software tasks developed by the AGILE
instrument teams and integrated into the  
Quick Look (QL) and standard analysis pipelines developed at ADC (Pittori et al. 2010).

AGILE is well suited for detection and alerts on $\gamma$-ray galactic and extra galactic transients, 
$\gamma$-ray bursts (GRB), X-ray bursts and other transients in the hard X-rays.
Both the mentioned QL pipelines running  at ADC and 
the AGILE Science Alert System developed and running at the AGILE Team Quick Look sites  
have been  designed to 
automate the identification of these astrophysical events in the AGILE data. 
In the following we present in particular the Alert System architecture and performances.

\vspace{.2cm} 

\smallskip

\begin{figure}[ht!]
\centerline{\psfig{figure=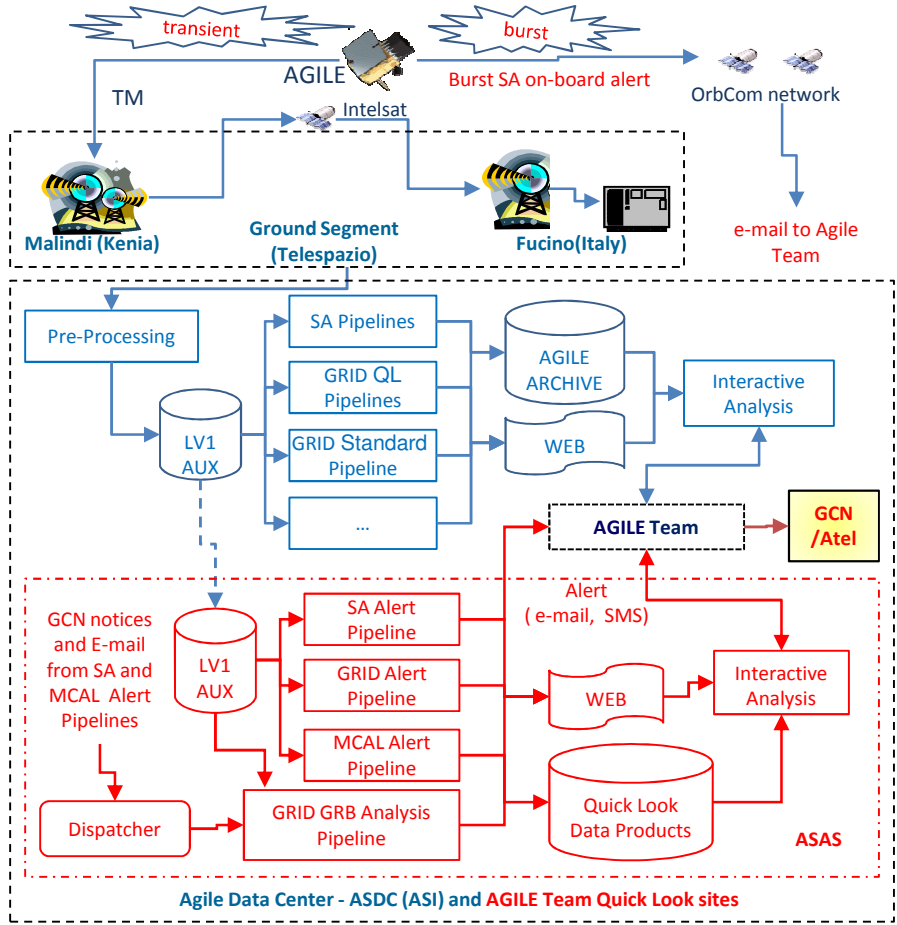,width=12cm}}
\caption{Overview of the AGILE Science Alert System  in the AGILE Ground Segment scenario.}
\end{figure}

\section{Alert System architecture and performances}

As  sketched in Figure 1, the Alert System consists of several tasks running at the AGILE Team Quick Look sites. 
The 3 Alert Pipelines (SA, GRID, MCAL) process the LV1 data as soon as they are received from ADC within {$\sim$}40 minutes after the TM down-link start ($T_0$) at the Ground Station, together with the auxiliary data (AUX) required for the AGILE orbit and attitude reconstruction.

The SA Alert Pipeline runs at IASF Roma to search for GRB, X-ray bursts and other transients in the hard X-ray band.
The GRID Alert Pipeline and the MCAL Alert Pipeline run at IASF Bologna. The former is devoted 
to detection and alerts on galactic and extra galactic transients in the Field of View (FOV) 
of the GRID detector. The latter processes the data of the MCAL burst chain in order to 
identify valid burst on-board triggers and to search for both Cosmic GRB and 
for valid Terrestrial Gamma-ray Flashes (TGF) candidates.
Each Pipeline, in case of transient or burst detection, alerts  the AGILE team via e-mail and via SMS 
to mobile phones.

In addition, at IASF Bologna the Dispatcher task handles the Swift/BAT, INTEGRAL/IBIS, and Fermi/GBM Notices received from the GRB Coordinates Network (GCN) through a TCP/IP Internet socket and the GRB alerts generated via e-mail by the above SA and MCAL Pipelines.
In case the GRB is localized inside the GRID Field of view (FoV), this task runs 
the GRID GRB Analysis Pipeline in order to search for a detection 
or provide an upper limit (UL) on the flux. The analysis results are e-mailed to the AGILE Team.

Alert System performance measurements have been collected over a period of 400 orbits in September 2009. 
They show that the time taken  by the Alert System  Pipelines to process the LV1 data of each 
orbit and generate the alert, if any, varies from {$\sim$}4-5 minutes, in the SA and MCAL 
cases, to  {$\sim$}38-74 minutes, in the GRID case. Hence, the time elapsed after $T_0$ is 
{$\sim$}44-45 minutes and {$\sim$}78-114 minutes, respectively. 

The Quick Look data products (e.g.: maps, light curves, spectra) are made available to the Team through 
Web pages. Local and remote access to the interactive analysis tools available at ADC and at Agile Team 
sites allow the Team to perform further analysis of the alert.
This has led to the issue of ATels within  $T_0$ + {$\sim$}0.5-1 day for both SA and GRID, and to the 
issue of GCNs within  $T_0$ + {$\sim$}1-3 hours for SA and MCAL, and $T_0$ + {$\sim$}12 hours for GRID. 
Futher details on the main Alert System components are given hereafter.

\subsection{SA Alert Pipeline}

The SuperAGILE ground software (Feroci et al. 2007) is equipped with a trigger algorithm in order to search for GRB, 
X-ray bursts and other transients in the hard X-ray band. The ground trigger runs on timescales from 512 ms up to 16384 ms 
using the Scientific Ratemeters telemetry data and taking advantage of the segmentation of the SuperAGILE detector.  
Up to now, the SuperAGILE ground trigger has detected several GRBs and X-ray bursts, including the remarkable flare 
from IGR J17473-2721 A remarkable flare from IGR J17473-2721 (26 March 2008) heralding the reactivation of the source 
and allowing its classification as Low Mass X-ray Binary.




\subsection{GRID Alert Pipeline}

This is an updated version of the AGILE GRID Science Monitoring system (Bulgarelli et al. 2009).
It consists of three main stages: (a) event
filtering, which applies the software modules of the AGILE standard analysis
(Vercellone et al. 2006) to reconstruct the event characteristics and filter out the
background events, (b) map generation, which executes the software modules
of the AGILE scientific analysis (Chen et al. 2009) to build maps of the last
24 hours of GRID observation, and (c) transient search to check for the presence of galactic or
extra-galactic transients in the FoV of the GRID detector. 

In the current version the search for candidate transient Galactic sources in daily AGILE GRID maps
is carried out with two statistical methods:
(i) blind-search (SPOT4) based on counts excesses
search and on the likelihood multi-source analysis; 
(ii) rate of false discoveries (FDR) in source detection,
that selects candidate sources with a FDR of $10^{-3}$.
The former uses as input the maps having FoV of 100$^o$. 
As first step (SpotFinder), it extracts from the intensity map the list of the more intensive spots.
Hence, it applies on these spots the likelihood multi-source analysis method.
The remaining operations are performed as in the previous mentioned version. 
FDR is a new functionality  that allows to
control the rate of false discoveries (FDR) in source detection (Sabatini et al. 2010). 
The search  is performed on counts maps and it is based on the idea that
the flux from a source stands on top of the background emission and it
is therefore expected to be on average higher than the background
flux. We can then define a threshold above which counts values have a
'low probability' to be statistical fluctuations of the background and
are therefore likely to be part of a source. In the use of this
thresholding technique the crucial point is the definition of the
threshold value, that one wants to be high enough to minimize the
contamination by false detections and low enough to select the higher
possible number of sources. The contamination and the completeness of
the selected sample of sources can be assessed statistically based on
the used threshold value. The FDR method allows to define the
threshold based on the set of data and controls the false discovery
rate in the case of multiple testing.
We developed an automatic analysis of the daily maps of AGILE GRID
that performs the FDR, selects candidate sources with a false
discovery rate of 10$^{-3}$ and sends an alert to the AGILE team.

\subsection{MCAL Alert Pipeline}

This is an updated version of the MCAL burst monitoring pipeline presented in Bulgarelli et al. 2008.
For every trigger detected by the on-board logic the event data are analyzed to exclude instrumental triggers. 
All valid bursts are then analyzed to extract automatically the main physical parameters (T90, fluency, peak flux). 
In 4-5 minutes the analysis completes, and the results as well as the light curve in different energy bands are sent to 
the Team by e-mail. In addition a suitable e-mail is sent to the Dispatcher in order to perform the burst search in the 
GRID (see 2.4).
Triggers on very short ($<$16ms) time scales are analyzed using different criteria to search for both Cosmic GRB and for 
valid Terrestrial Gamma-ray Flashes (TGF) candidates. 
Information on good TGF candidates are notified automatically to Agile Team and external collaborators for rapid correlation 
with atmospheric (Very-Low Frequency - sferics ) data. 
Information on all on-board triggers are stored in a database that can be accessed from the WEB for monitoring. 
In addition, on contact by contact basis, a ground trigger algorithm runs on scientific ratemeters to notify via e-mail the 
MCAL team for transients not recognized by the on-board logic. 

\subsection{GRID GRB Analysis Pipeline}

As mentioned above, an Alert System task handles the Notices received from the GCN. 
For each Notice, it waits the AGILE telemetry to estimate the visibility of the burst position at the trigger time.
In case the GRB is localized inside the GRID FoV, it sends, through the Dispatcher, the burst information (time start, RA, DEC) 
to the Pipeline. A map showing the GRB direction and the AGILE pointing is produced
and mailed to the Team.
The same run modality (i.e. with coordinates ) applies for the e-mails generated by the SA Alert Pipeline. 
The GRID GRB Analysis Pipeline performs the GRB search in the AGILE data by comparing the expected
background with the counts in the first 60s in a region of 15$^o$ of radius centered
at the position of the burst. The method used to estimate the detection and the UL (Moretti et al. 2009)
is based on the derivation of the Bayes theorem. A good estimation
of the background is required to use this approach, to know with negligible error
the mean background rate. The background mean is calculated in a time interval at
least 10 times longer the duration of burst and before the trigger. During this time
interval both background and signal are supposed to follow a Poisson distribution
and the signal to be non-negative.

\end{document}